# Probing Electronic States in Monolayer Semiconductors through Static and Transient Third-Harmonic Spectroscopy

*Yadong Wang\*, Fadil Iyikanat, Habib Rostami, Xueyin Bai, Xuerong Hu, Susobhan Das, Yunyun Dai, Luojun Du, Yi Zhang, Shisheng Li, Harri Lipsanen, F. Javier García de Abajo[*], and Zhipei Sun[*]*

Y. Wang, X. Bai, X. Hu, S. Das, Y. Dai, L. Du, Y. Zhang, H. Lipsanen, Z. Sun

Department of Electronics and Nanoengineering, Aalto University, Espoo 02150, Finland

E-mail: yadong.wang@aalto.fi

E-mail: zhipei.sun@aalto.fi

F. Iyikanat, F. J. García de Abajo

ICFO-Institut de Ciencies Fotoniques, The Barcelona Institute of Science and Technology, 08860 Castelldefels (Barcelona), Spain

H. Rostami

Nordita, KTH Royal Institute of Technology and Stockholm University, Hannes Alfvéns väg 12, 10691 Stockholm, Sweden

S. Li

International Center for Young Scientists, National Institute for Materials Science, Tsukuba, Japan

F. J. García de Abajo

ICREA-Institució Catalana de Recerca i Estudis Avançats, Passeig Lluís Companys 23, 08010 Barcelona, Spain

E-mail: javier.garciadeabajo@nanophotonics.es

Z. Sun

QTF Centre of Excellence, Department of Applied Physics, Aalto University, Espoo, Finland

E-mail: zhipei.sun@aalto.fi

**Abstract:** Electronic states and their dynamics are of critical importance for electronic and optoelectronic applications. Here, we probe various relevant electronic states in monolayer $MoS_2$, such as multiple excitonic Rydberg states and free-particle energy bands, with a high relative contrast of up to ≳200 via broadband (from ~1.79 to 3.10 eV) static third-harmonic spectroscopy, which is further supported by theoretical calculations. Moreover, we introduce transient third-harmonic spectroscopy to demonstrate that third-harmonic generation can be all-optically modulated with a modulation depth exceeding ~94% at ~2.18 eV, providing direct evidence of dominant carrier relaxation processes, associated with carrier-exciton and carrier-phonon interactions. Our results indicate that static and transient third-harmonic spectroscopies are not only promising techniques for the characterization of monolayer semiconductors and their heterostructures, but also a potential platform for disruptive photonic and optoelectronic applications, including all-optical modulation and imaging.

## 1. Introduction

Two-dimensional (2D) layered materials such as transition metal dichalcogenides (TMDs) have emerged as promising candidates for advanced electronic and optoelectronic applications due to their unique physical properties.[1] Assisted by the intrinsic strong Coulomb interaction, many-particle electronic states (e.g., Rydberg excitons) in monolayer TMDs produce pronounced electronic and optical responses.[2] In order to fulfil their potential applications, it is vital to fully characterize and understand such TMD electronic states, as well as their dynamics.[3-11] In this direction, the *s*-series of excitons allowed in the electric-dipole (ED) approximation have been investigated using various linear optical spectroscopies, such as differential linear optical reflection,[12] photoluminescence[13] and photocurrent spectroscopies,[14] while higher-energy states remain challenging to access due to their intrinsic relatively-weak light-matter interaction.

Importantly, nonlinear light interactions[15-22] (such as harmonic generation) usually provide a high signal contrast. For example, a signal variation up to three orders of magnitude has been demonstrated through second-harmonic spectroscopy.[19, 22] In addition, third-harmonic generation (THG) is allowed in all media regardless of their lattice symmetry. Thus far, third-harmonic spectroscopy (THS) has been widely utilized in the exploration of physical features displayed by various materials, such as semiconductors, metals, and bio-materials.[23] In particular, THG in monolayer TMDs has been observed to reach significantly stronger levels than SHG at the same excitation photon energy[24-26], let alone that in other centrosymmetric monolayers, such as graphene[27-31] and black phosphorus[32-34]. Such strong THG could arise from resonances of electronic states, which offer an opportunity to probe the electronic states with high contrast.[23] However, detailed investigation of THG in monolayer TMDs responding to the resonant states still remains to be performed.[35, 36]

Here, we demonstrate broadband (~1.79 to 3.10 eV) THS in monolayer $MoS_2$ to explore prominent fingerprints of its optically relevant electronic states. This approach grants us direct access into the *s* series of excitons in monolayer TMDs via multi-photon resonance processes. The THG efficiency not only sensitively depends on the exciton states, but also exhibits a strong correlation with the free-band electronic transitions, which we confirm through first-principles theoretical calculations. The carrier dynamics at various states are further characterized by transient THS, providing strong signatures of exciton-induced THG with carrier trajectories involving carrier-exciton interactions and carrier-phonon interaction in monolayer $MoS_2$. Our demonstration of THS in monolayer MoS$_2$ allows us to establish a detailed picture of electronic states in monolayer TMD, including *s* series excitons and free-band electronic states.

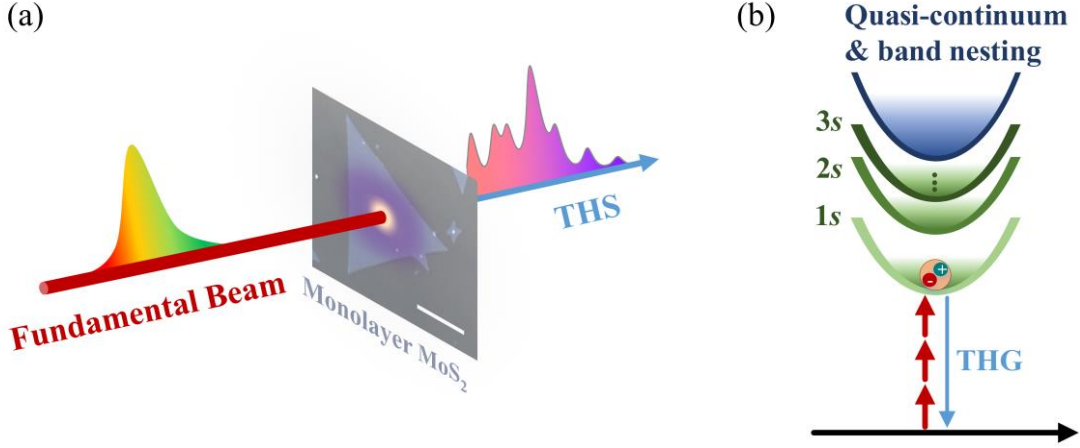

**Figure 1. THS of monolayer MoS$_2$: Experimental configuration and emission mechanism.** (a) Simplified scheme of THG in monolayer MoS$_2$. A broadband-tunable fundamental seed laser is used to produce THG upon normal incidence on monolayer MoS$_2$. The scale bar in the optical image is 20 μm. (b) Schematic of three-photon resonant THG.

## 2. Results and discussions
### 2.1 Broadband THS in monolayer MoS$_2$

We study the THS in monolayer MoS$_2$ (optical band gap: ~1.85 eV [37]) flakes grown through chemical-vapor-deposition (CVD) as an instance of TMDs. An optical image of the MoS$_2$ flake with a typical triangular structure on a sapphire substrate is shown in **Figure 1(a)**. Several optical characterization methods (such as linear reflectance, Raman, and photoluminescence spectroscopies) are used to confirm the high quality of our monolayer MoS$_2$ flakes (see details in Section 1 of the Supplemental Information). In order to carry out THS on monolayer MoS$_2$, we use ~150 fs incident fundamental pulses with photon energies ($\hbar\omega_0$) tunable from ~0.60 to 1.03 eV (wavelength $\lambda_0$: ~1200 to 2080 nm) to generate third-harmonic pulses (Figure 1(a)) with photon energy ($\hbar\omega_{THG}$) in the range of ~1.79 to 3.10 eV (the corresponding wavelength ($\lambda_{THG}$) in the range of ~400 to 693 nm, see details in Section 2 of the Supplemental Information). The detailed THG characterization setup is shown in Section 3 of the Supplemental Information. A scheme of possible resonance conditions is presented in Figure 1(b).

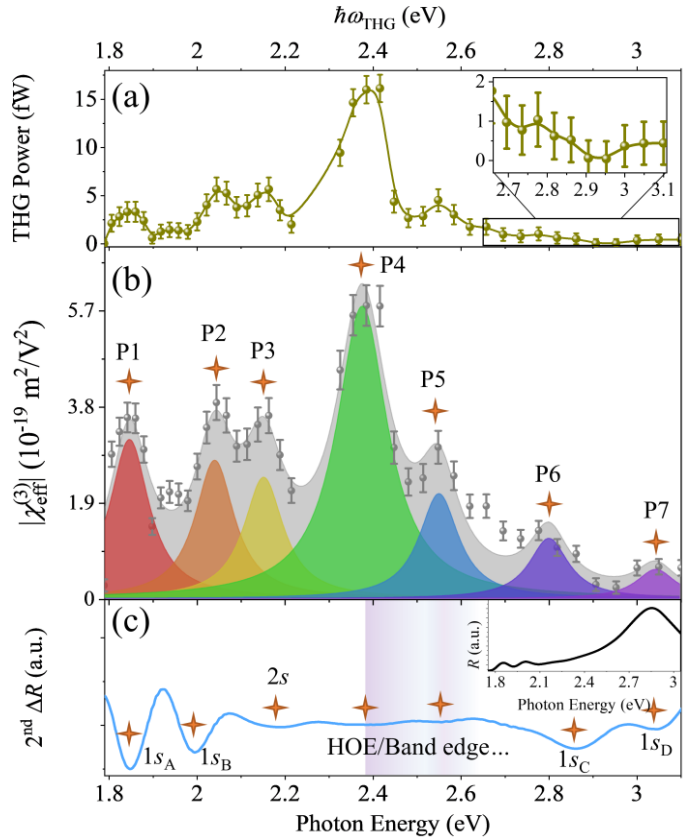

**Figure 2. THS of monolayer MoS$_2$: Experimental results.** (a) THG power and (b) experimentally evaluated $|\chi^{(3)}_{\text{eff}}|$ as a function of THG photon energy, along with (c) the second-order linear reflectance (2$^{\text{nd}}$ $\Delta R$) from monolayer MoS$_2$. Inset to (a): zoom of the ~2.67-3.10 eV region. Inset to (c): linear reflectance of monolayer MoS$_2$. Series peaks in (b) are labelled as P1-P7, with their maxima indicated by stars. Series electronic states (including excitonic states) are labelled in (c).

Figure 2(a) shows THG results at different output light wavelengths in monolayer MoS$_2$ with a fixed incident average power ~10 μW (peak power: ~141 GW/cm$^2$). The results reveal a strong dependence of the THG signal on the incident fundamental wavelength. The variation of the THG intensities can reach a factor of ~800 between ~1.79 eV and 3.10 eV, accompanied by remarkable THG enhancement at certain photon energies. The signal contrasts at different selected energies are given in Table S2 of the Supplemental Information, in which most of them reach high values (≳200). The measurements carried on different flakes show that our THS is reliable to probe the series of electronic states in monolayer TMDs. The THG ($\hbar\omega_{\text{THG}} \approx 2.38$ eV) efficiency is the highest when $\hbar\omega_0$ is ~0.79 eV ($\lambda_0 \approx 1560$ nm). We then calculate the effective third-order nonlinear susceptibility $|\chi_{\text{eff}}^{(3)}|$ at different THG energies (grey dots in Figure 2(b)). Details about the calculation are provided in Section 3 of the Supplemental Information. In Figure 2(b), seven resonant peaks (P1-P7) appear to be well fitted with Lorentzian functions at output photon energies (from P1 to P7) ~1.84, 2.03, 2.16, 2.38, 2.55, 2.81, and 3.05 eV, respectively. Out of these, P4 is the strongest with $|\chi_{\text{eff}}^{(3)}|$ of ~5.7×10$^{-19}$ m$^2$/V$^2$, which is comparable to previous results for a similar wavelength region.[24, 25, 38]

To understand the emerging multiple peaks in the THS measurements, we present the second-order contrast derivative (2$^{\text{nd}}$-$\Delta R$) of the linear reflection spectrum of monolayer MoS$_2$ in Figure 2(c). We use the 2$^{\text{nd}}$-$\Delta R$ because it renders a better contrast with respect to excitonic positions[12] than the original reflection (see inset to Figure 2(c)). Indeed, seven dips are observed in Figure 2(c), with photon energies of ~1.85, 2.00, 2.17, 2.38, 2.55, 2.84, and 3.05 eV, respectively. Interestingly, the THS peaks nicely match the dips in the 2$^{\text{nd}}$-$\Delta R$ spectrum, suggesting the involvement of resonant three-photon excitation at electronic states. Among them, three dips in Figure 2(c) at ~1.85 eV, 2.00 eV, and 2.84 eV clearly correspond to the A, B, and C excitons with 1$s$ series, matching the THS peaks at P1, P2, and P6 in Figure 2(b), respectively (i.e., excitations from the ground state to the indicated excitons). This is similar to previously reported exciton-enhanced SHG results in monolayer TMDs.[19, 39] In addition, the dip at ~2.17 eV in Figure 2(c) (and P3 in Figure 2(b)) could be assigned to the 2$s$ exciton,[13] while the dip at ~2.38 eV in Figure 2(c) (and P4 in Figure 2(b)) could possibly correspond to the 3$s$ or higher-order excitons (HOEs).[13] We discuss more details about P4 below. In contrast, P5 in Figure 2(b) and the corresponding weak dip at 2.55 eV in Figure 2(c) cannot easily be ascribed and we discuss it later as well. Finally, the well-known band-nest effect contributes to the linear absorption features at ~2.84 (1$s_\text{C}$) and 3.05 eV (1$s_\text{D}$), matching the peaks P6 and P7 in Figure 2(b). Importantly, all 7 electronic states of monolayer MoS$_2$ are nicely producing a matching between the 2$^{\text{nd}}$-$\Delta R$ spectrum and THS (see Table I), showing that static THS provides an alternative method to detect optically relevant carrier states, but with higher contrast of up to ≳200.

To further support the above analysis, we carry out first-principles calculations of the electronic structure, including excitonic effects. Specifically, we use many-body perturbation theory (MBPT) and solve the Bethe-Salpeter equation on top of the G$_0$W$_0$ electronic structure, followed by real-time simulations to describe nonlinear optical effects. These methods have been shown to yield reliable estimates that are highly consistent with experimental results.[40, 41] In Figure 3(a), we show the absolute value of $\chi_{\text{eff}}^{(3)}$ calculated for monolayer MoS$_2$ within different levels of theory, namely, the independent-particle approximation (G$_0$W$_0$-IPA) and the Bethe-Salpeter equation (G$_0$W$_0$-BSE) on top of G$_0$W$_0$. The THG spectrum calculated using the G$_0$W$_0$-IPA model in Figure 3(a) clearly shows the THG processes originating from the band edge. In the calculations in which the excitonic effects are included by solving the Bethe-Salpeter equation, new prominent peaks emerge at energies much lower than the electronic band gap, in agreement with experimental observations. Therefore, it is clear that peaks at low energies have their origin in the excitons, whereas the strong and broad peak observed at the electronic band edge is due to both band-edge transitions and excitonic transitions at those energies. The calculated peak positions of the excitons with high oscillator strength are given in Table S1, along with the electronic band gap.

The results from the G$_0$W$_0$-IPA approach show the electronic band gap at ~2.73 eV emerging as a shoulder of the main profile. It also shows two main peaks at ~2.83 and 2.99 eV above the electronic band edge. In contrast, the inclusion of excitonic effects inherent in the G$_0$W$_0$-BSE approach gives rise to dramatic changes in the THG spectrum, which now exhibits four main exciton-originated excitations located below the electronic band edge at THG output photon energies of ~2.04, 2.18, 2.40, and 2.70 eV. Clearly, the first two peaks located at 2.04 and 2.18 eV, corresponding to $1s_A$ and $1s_B$ excitons, show the highest THG signal near the optical band edge with a magnitude $|\chi_{\text{eff}}^{(3)}| \simeq 4 \times 10^{-18}$ m$^2$/V$^2$. The other prominent and rather broad optical transition peak is located at 2.70 eV and arises from contributions of the HOE/Band edge induced transitions and $1s_C$ exciton, yielding a THG magnitude $|\chi_{\text{eff}}^{(3)}| \simeq 2.9 \times 10^{-18}$ m$^2$/V$^2$. By reducing the damping parameter used in the calculations, it is also possible to reveal additional peaks appearing in the THG spectrum, indicating the involvement of higher-order excitons. The difference between the experimentally measured and calculated band gap and peak positions can be attributed to the inadequacy of the parameter sets used in the calculations, since the computational cost of THG calculations is quite high. To prove this, we also plot the imaginary part of the dielectric function of monolayer MoS$_2$ calculated with different sets of parameters in Figure S6(a) of the Supplemental Information, which reveals that, as the parameter set used in the calculation of the dielectric function is improved, the obtained band gap values and peak positions approach the experimentally observed values. The direct electronic band gap at the K point of the Brillouin zone is calculated to be $E_g$ ~2.73 eV (and 2.48 eV with an optimized set). The agreement between the prominent peak positions experimentally observed in the THG spectrum (Figure 2(a), (b)) and the calculated exciton peak positions (Figure 3(a), (b)) in the dielectric function is clear. Measured and calculated peak positions of excitons with prominent contributions to the THG spectra and a scheme of the exciton-induced THG process are schematically illustrated in Figure 3(c).

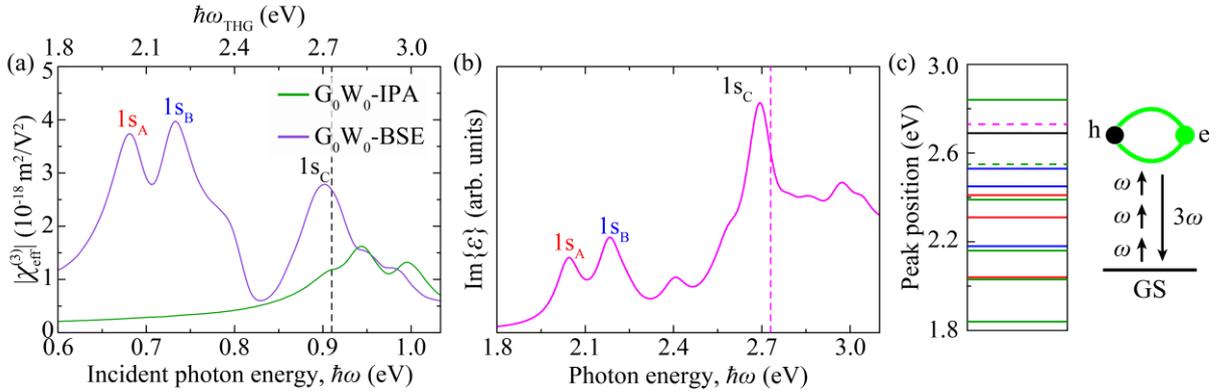

**Figure 3. First-principles theory of THS in monolayer MoS₂.** (a) Nonlinear THG susceptibility $|\chi_{\text{eff}}^{(3)}|$ as a function of incident photon energy, calculated within the independent-particle approximation (IPA) and Bethe-Salpeter equation (BSE) methods on top of the $G_0W_0$ electronic structure. The vertical dashed line indicates the calculated band gap. (b) Imaginary part of the dielectric function and electronic band gap (vertical dashed line) of monolayer MoS₂ calculated with the $G_0W_0$-BSE method. (c) Sketch of the relevant exciton energy levels located near the optical band edge and having high oscillator strengths. The red and blue lines represent the calculated spectral position (1$s$, 2$s$, and 3$s$, in order of increasing energy) of A and B excitons, respectively, while the black line shows the 1$s_C$ exciton. Green lines indicate the peak positions observed in the THG measurements. The purple and green dashed lines denote the calculated and measured band gap, respectively. The right scheme illustrates the process of THG associated with three-photon excitation from the ground state to an excitonic state.

After understanding the exciton-induced part of THG, two concerns are remaining. One is that the peak P4 at ~2.38 eV in Figure 2(b) is extremely strong, in contrast to the weaker feature associated with the high-order exciton states.[12] Also, the fitted FWHM width of peak P4 in Figure 2(b) is ~0.14 eV, larger than the other six peaks, which have a typical FWHM of ~0.11 eV. Such substantial enhancement and extra broadening possibly indicate the contribution of additional resonances to peak P4, besides the 3$s$ excitonic state. This is consistent with our $G_0W_0$-BSE-based calculations, revealing that the peak near the electronic band edge is a combination of several HOEs such as 3$s$ and 4$s$ excitonic states. Thus, we assign the P4 feature as the HOE/band edge. In addition, we find that the energy difference between P4 and P5 (~0.17 eV) is close to that between $1s_A$ and $1s_B$ excitons (~0.15 eV), therefore pointing to the band edge between the A and B series, so that P4 is tentatively denoted as HOE/Band(A) and P5 could be accordingly assigned as the band edge (B) (Band(B)). However, a conclusive assignment of P4 and P5 requires future investigation. Incidentally, we observe that the enhancement produced by band nesting (P6 and P7) on THG is one order of magnitude smaller than that on linear and SHG responses.[39] We further calculate the value of nonlinear transition matrix elements and find that the impact of density of states at the band-nesting region on the THG intensity is suppressed by the small value of the nonlinear transition matrix elements, which indicates that weak THG at $1s_C$ and $1s_D$ excitons (See Section 6 in the Supplemental Information).

## 2.2 Ultrafast transient THS in monolayer MoS$_2$

We carry out transient THS to study the carrier dynamics and to further understand the THG enhancement, which is of vital importance for electronic and optoelectronic applications. The left panel of Figure 4(a) shows a diagram of transient THS. With the excitation of pump pulses at higher photon energy, carriers are excited to higher states and then relax to lower states (right panel of Figure 4(a)), therefore giving rise to modulation of THG and offering a good way to identify the relevant electronic states and mechanisms of relaxation dynamics. In our experiment, a pump light beam at ~3.1 eV is used with a delay line to adjust the time delay ($\Delta\tau$) between the pump light and a wavelength-tunable seed light (i.e., the fundamental excitation pulses for THG) (see Section 2 in the Supplemental Information).

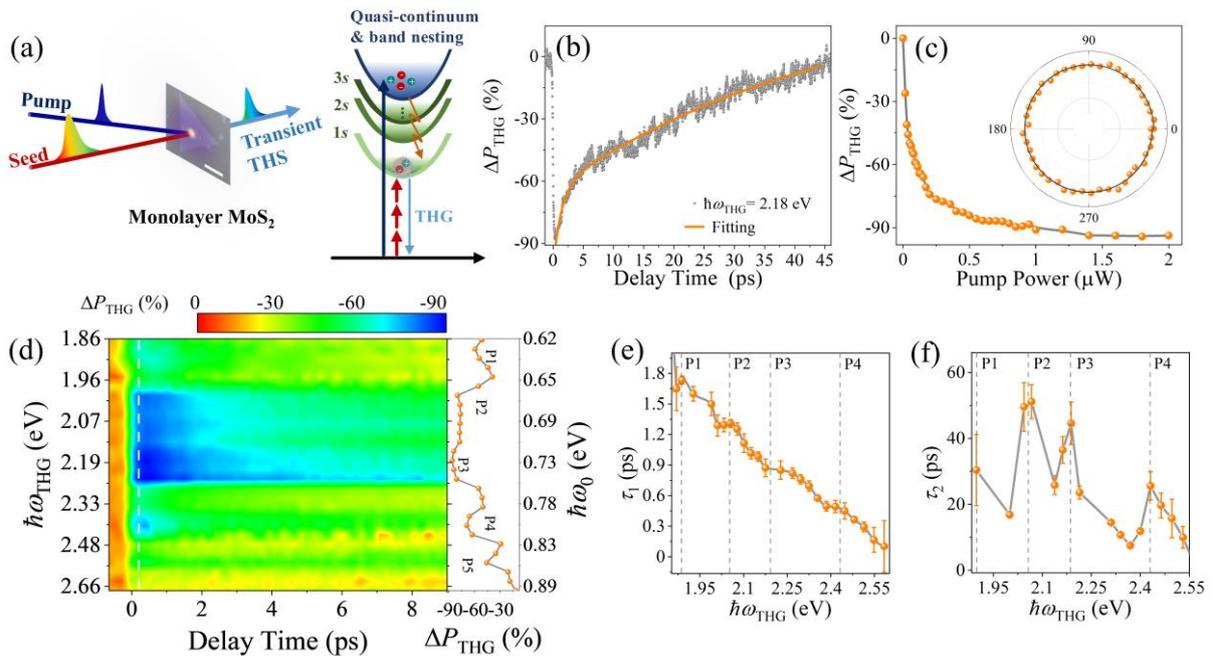

**Figure 4. Transient THS of monolayer MoS$_2$.** (a) Schematics of transient THS. Left panel: illustration of transient THS; right panel: illustration of carrier relaxation during the transient THG process involving different electronic states. (b) Transient fractional change of THG emission $\Delta P_{THG}$ when the seed and pump light powers are ~10 and 1 μW, respectively. The decay process is fitted by a bi-exponential function (yellow curve) with a fast time constant of ~1.3 ps and a slow time constant of ~46.4 ps. (c) $\Delta P_{THG}$ as a function of pump light power. (d) Transient THS at different output photon energies $\hbar\omega_{THG}$ from ~1.86 to 2.68 eV. The right panel shows $\Delta P_{THG}$ at different photon energies for a fixed time delay of ~0.3 ps. (e) Fast $\tau_1$ and (f) slow $\tau_2$ decay time constants as a function of THG photon energy. The vertical dashed lines in (e) and (f) indicate the positions of P1-P4 peaks, while the grey lines in (c)-(f) are guides to the eye.

Figure 4(b) shows the transient THG signal ($\Delta P_{THG}$) at the generated photon energy of ~2.18 eV. $\Delta P_{THG}$ is defined as the fractional change in THG power by Equation $\Delta P_{THG}=(P_\omega - P_0)/P_\omega$, where $P_\omega$ is the THG intensity in the presence of the pump excitation and $P_0$ is the THG intensity without the pump excitation. In Figure 4(b), the observed THG intensity decreases quasi-instantaneously within ~100 fs after the pump pulses arrive, and it eventually reaches its lowest value of -88% at $\Delta\tau$ = ~0.3 ps. In the measurements, pump excitation of charge carriers bleaches the corresponding multi-photon transitions underlying the THG process, thus the THG efficiency is correspondingly reduced.[42] After, the THG signal begins to recover with a typical bi-exponential trajectory, whose fitted time constants are $\tau_1$ = ~1.3 ps and $\tau_2$ = ~46.4 ps for the fast- and slow-time components. This indicates the presence of two types of processes of carrier dynamics in the crystal lattice[43], similar to previous studies: i) The fast carrier relaxation time is mainly associated with carrier-carrier relaxation. This process usually happens within a few picoseconds.[10, 44, 45] ii) The slow carrier relaxation time is mainly due to carrier-phonon interactions. Note that the additional interactions of carriers with defects can produce charge trapping and thus also possibly contributes to a slow relaxation time.[46, 47] By fixing the delay time ($\Delta\tau$) at ~0.3 ps, we observe that $\Delta P_{THG}$ continuously decreases to -94% as a function of the pump power until the average pump power reaches ~1.8 μW (Figure 4(c)). This nonlinear behavior of the power dependence could be attributed to complete carrier depletion. Incidentally, the inset of Figure 4(c) shows $\Delta P_{THG}$ as a function of the pump light polarization, indicating that the THG modulation is independent of the pump polarization.

To find the relationship between electronic states and $\Delta P_{THG}$, we collect measurements of $\Delta P_{THG}$ by scanning the excitation seed energy from ~0.62 to ~0.89 eV (at fixed average seed light power: 10 µW) with ~1-µW pump pulses at ~3.1 eV (Figure 4(d)). The results show that the THG intensity decreases in the whole spectral region. Specifically, when we extract the $\Delta P_{THG}$ response at a delay time $\Delta\tau =$ ~0.3 ps, as shown in the right panel of Figure 4(d), it clearly shows five dips at ~1.87, 2.00, 2.18, 2.40, 2.55 eV with significant modulation depths (see details in Table S2 of the Supplemental Information), respectively, which agree well with the THS results (i.e., peaks P1-P5 in Figure 2(b)) and the reflectance spectrum in Figure 2(c) (see Table I). Incidentally, $\Delta P_{THG}$ at the C exciton (~2.8 eV) does not exhibit any feature due to the weak THG signal. We find that the 2$s$-state-enhanced THG at peak P3 enables the highest modulation depth, while that associated with HOE/Band(A) at peak P4 is relatively weak, which shows that the THG modulation at the band edge is less efficient than that arising from exciton states. This indicates a contribution of the electronic band gap to the THG enhancement at peak P4, which adds up to the contribution originating in the higher-level excitonic states, in agreement with our theoretical calculation (see above discussion of Figure 3).

We plot the fitted decay time constants as a function of the THG photon energy in Figure 4(e) and 4(f). The fast time constant $\tau_1$ decreases with the THG photon energy from ~2.47 ps at ~1.88 eV to ~0.1 ps at ~2.64 eV. This is reasonable as it takes a longer time for carriers to relax at the lower energy states.[48] In addition, we observe several ladders marked with P1-P4, which could be attributed to the carrier-exciton and carrier-phonon interactions.[10, 49] The slow time constant $\tau_2$ increases when it is close to the resonant states (i.e., 1$s_A$ (P1, ~1.89 eV), 1$s_B$ (P2, ~2.05 eV), 2$s$ (P3, ~2.18 eV), HOE/Band(A) (P4, ~2.43 eV)), as shown in Figure 4(f), indicating that the bounded exciton states can also affect the recombination process.[50] With transient THS, carriers can be monitored with high modulation depth (~94% at ~2.18 eV in Figure 4(c)) and high time resolution (~0.15 ps in Figure 4(b), 4(e)), providing an accurate method to study carrier relaxation processes. This method could also be applied to other TMDs and heterostructures to explore the dynamics of interlayer electronic states, such as moiré excitons.[51, 52] Further, such all-optical nonlinear modulation with high modulation depth and ultrafast speed could enable the development of emerging all-optical nonlinear photonic applications.[53-56]

**Table I** Comparison of electronic states obtained with different methods

| Characterization methods | $1s_A$ (eV) | $1s_B$ | $2s$ | HOE/Band(A) | Band(B) | $1s_C$ | $1s_D$ |
|---|---|---|---|---|---|---|---|
| $2^{nd}\,\Delta R$ | 1.85 | 2.00 | 2.17 | 2.38 | 2.55 | 2.84 | 3.05 |
| THG | 1.84 | 2.03 | 2.16 | 2.38 | 2.55 | 2.81 | 3.05 |
| Transient THG | 1.87 | 2.00 | 2.18 | 2.40 | 2.55 | * | * |
| $G_0W_0$-BSE calculations (48×48, 300) | 1.89 | 2.03 | 2.19 | 2.42 | 2.48 | 2.69 | * |

\* not available

## 3. Conclusion

We have introduced broadband and high-contrast static and transient THS as new characterization methods, and applied them to monolayer semiconductor to uncover several relevant electronic states such as *s* series excitons and interlayer band-to-band transition effects. The new technique offers a unique way of probing such states, with experimental results admitting a direct interpretation, as supported by comparison with first-principles theory, in which the THG behavior is described through the real-time evolution of Bloch electrons exposed to a time-dependent electric field. Furthermore, carrier dynamics in monolayer $MoS_2$ have been neatly observed through highly sensitive transient THS with a modulation depth reaching up to 94% at ~2.18 eV. By combining static and transient THS, we have demonstrated an efficient way to probe electronic and excitonic states as well as their dynamics. In addition, with such ultrafast and high modulation depth of transient THS, nonlinear optical modulation driven by excited electronic states holds great potential for nonlinear signal engineering.

**Supporting Information**

Supporting Information is available from the Wiley Online Library or from the author.

**Acknowledgements**

Yadong Wang and Fadil Iyikanat contributed equally to this work. The authors acknowledge the financial support from Aalto Centre for Quantum Engineering, Academy of Finland (grants: 314810, 333982, 336144 and 336818), Academy of Finland Flagship Programme (320167, PREIN), the European Union's Horizon 2020 research and innovation program (Grant agreement No. 820423,S2QUIP; 965124, FEMTOCHIP), Foundation for Aalto University Science and Technology and Finnish Foundation for Technology Promotion (grant: 8216), the EU H2020-MSCA-RISE-872049 (IPN-Bio), and ERC (834742).

**Conflict of Interest**

The authors declare no conflict of interest.